\documentclass[a4paper]{jpconf}
\usepackage{graphicx}
\begin{document}
\title{Central regions of LIRGs: rings, hidden starbursts, Supernovae and star clusters}

\author{Petri V\"ais\"anen$^1$, Andres Escala$^2$, Erkki Kankare$^3$, Jari Kotilainen$^4$, Seppo Mattila$^3$, Vinesh Rajpaul$^5$, Zara Randriamanakoto$^{1,5}$, Juha Reunanen$^3$, Stuart Ryder$^6$, Albert Zijlstra$^7$}

\address{$^1$South African Astronomical Observatory,  P.O. Box 9 Observatory 7935, South Africa, \\
$^2$ Departamento de Astronomia, Universidad de Chile, Casilla 36-D, Santiago, Chile\\
$^3$ Tuorla Observatory, University of Turku, FI-21500 Piikki\"o, Finland\\
$^4$ FINCA, University of Turku, V\"ais\"al\"antie 20, FI-21500 Piikki\"o, Finland\\
$^5$ Department of Astronomy, University of Cape Town, Rondebosch 7701, South Africa\\
$^6$ Anglo-Australian Observatory, Epping, NSW 1710, Australia\\
$^7$Jodrell Bank Centre for Astrophysics, University of Manchester, Manchester M13 9PL, UK\\

}

\ead{petri@saao.ac.za}

\begin{abstract}
We study star formation (SF) in very active environments, in luminous IR galaxies, which are often
interacting.  A variety of phenomena are detected, such as central starbursts, circumnuclear SF, obscured SNe tracing the history of recent SF, massive super star clusters, and sites of strong off-nuclear SF.  All of these can be
ultimately used to define the sequence of triggering and propagation of star-formation
and interplay with nuclear activity in the lives of gas rich galaxy interactions and mergers.  
In this paper we  present analysis of high-spatial resolution integral field spectroscopy of central regions of two interacting LIRGs.  We detect a nuclear 3.3 $\mu$m PAH 
ring around the core of NGC~1614 with thermal-IR IFU observations.  The ring's characteristics and relation to the strong star-forming ring detected in recombination lines are presented, as well as a scenario of an outward expanding starburst likely initiated with a (minor) companion detected within a tidal feature.  We then present NIR IFU observations of IRAS~19115-2124, aka the Bird, which is an intriguing triple encounter.   The third
component is a minor one, but, nevertheless, is the source of 3/4 of the SFR of the whole system. 
Gas inflows and outflows are detected at the locations of the nuclei.  
Finally, we briefly report on our on-going NIR adaptive optics imaging survey of several 
dozen LIRGs.  We have detected highly obscured core-collapse SNe in the central kpc, and 
discuss the statistics of "missing SNe" due to dust extinction.  We are also determining the 
characteristics of hundreds of super star clusters in and around the core regions of LIRGs, as a
 function of host-galaxy properties. 
\end{abstract}

\section{Introduction}

Luminous Infrared Galaxies, LIRGs, defined as galaxies with $10^{11} >  L_{IR} / L_{\odot} > 10^{12}$,
combine very strong star-formation (SF) with various levels of nuclear activity.  They are usually
more diverse objects regarding morphology and mode of SF than ULIRGs ($L_{IR}  > 10^{12} L_{\odot}$), which in the Local Universe tend to nearly always be remnants of major mergers with singularly 
centralised SF.   Some LIRGs also have centralised SF, but in many cases a significant part of the SF is ``extended''  (e.g.\ Alonso-Herrero et al. 2006; Rodr\'iguez-Zaur\'in et al. 2011).   It appears that high-redshift ULIRGs also have a  diversity of modes of strong SF present and hence the local LIRGs may be a closer representation of the physical sites of dominant SF density at $z \sim 1-2$ and beyond(e.g. Alonso-Herrero et al. 2009; Rodighiero et al. 2011; Kartaltepe et al. 2011).


"Extended SF" is often defined as merely a fraction of MIR or hydrogen recombination line emission coming from outside a given nuclear region, and as such it could mean either truly diffuse SF, or rather one or more widely off-nuclear, but clearly localised starbursts.  This latter scenario is playing out for example in the well-studied closest gas-rich merger system, the Antennae, where the SF is dominated by an overlap region between the main nuclei. To study the physics of such cases in detail, including disentangling various kinematic components,  requires high spatial resolution, which has not often been available except for a handful of local galaxies, .

We are conducting an ongoing large survey of LIRGs and ULIRGs using adaptive optics NIR
imaging mainly  with VLT/NACO, and also Gemini/ALTAIR/NIRI. Both instruments deliver images with a spatial resolution of $\approx 0.1$''. The data are very well complemented with existing HST optical data for many of the targets, and there is also an on-going optical spectroscopic campaign with SALT/RSS.  The galaxies are at distances ranging from 40 to 180 Mpc, span the range from starbursts to LIRGs, and are at various stages of merging, interaction, or isolation.  Some galaxies have been studied in more detail already -- here we will present analysis of two such objects, and will finish with a brief discussion of some initial results from the larger sample.

\section{NGC 1614: PAH rings and a minor companion}

NGC~1614 is a LIRG at a distance of 64 Mpc ($z=0.016$).   Optically it is classified as 
barred and interacting (Fig.~1), and  as a borderline case between a LINER and HII-dominated galaxy.  
It has also been classified as a composite starburst + AGN.   Previous studies have
identified a strong nuclear star forming ring of approximately 600 pc diameter (Kotilainen et al 2001; Alonso-Herrero et al. 2001).   
We observed the central 1.5 kpc region with UKIRT/UIST integral field unit in the L-band (V\"ais\"anen et al. 2012).
This wavelength region has been shown to be ideal for disentangling AGN signatures from 
starbursts, using both the continuum slope and the 3.3 and 3.4 $\mu$m PAH band features 
(e.g.\ Imanishi \& Dudley 2000; Risaliti et al. 2006).
The method has not, however, been much used with both spatial and spectroscopic information simultaneously, i.e.\ with IFU, due to the rarity of suitable instrumentation.

\begin{figure}[h]
\includegraphics[width=36pc]{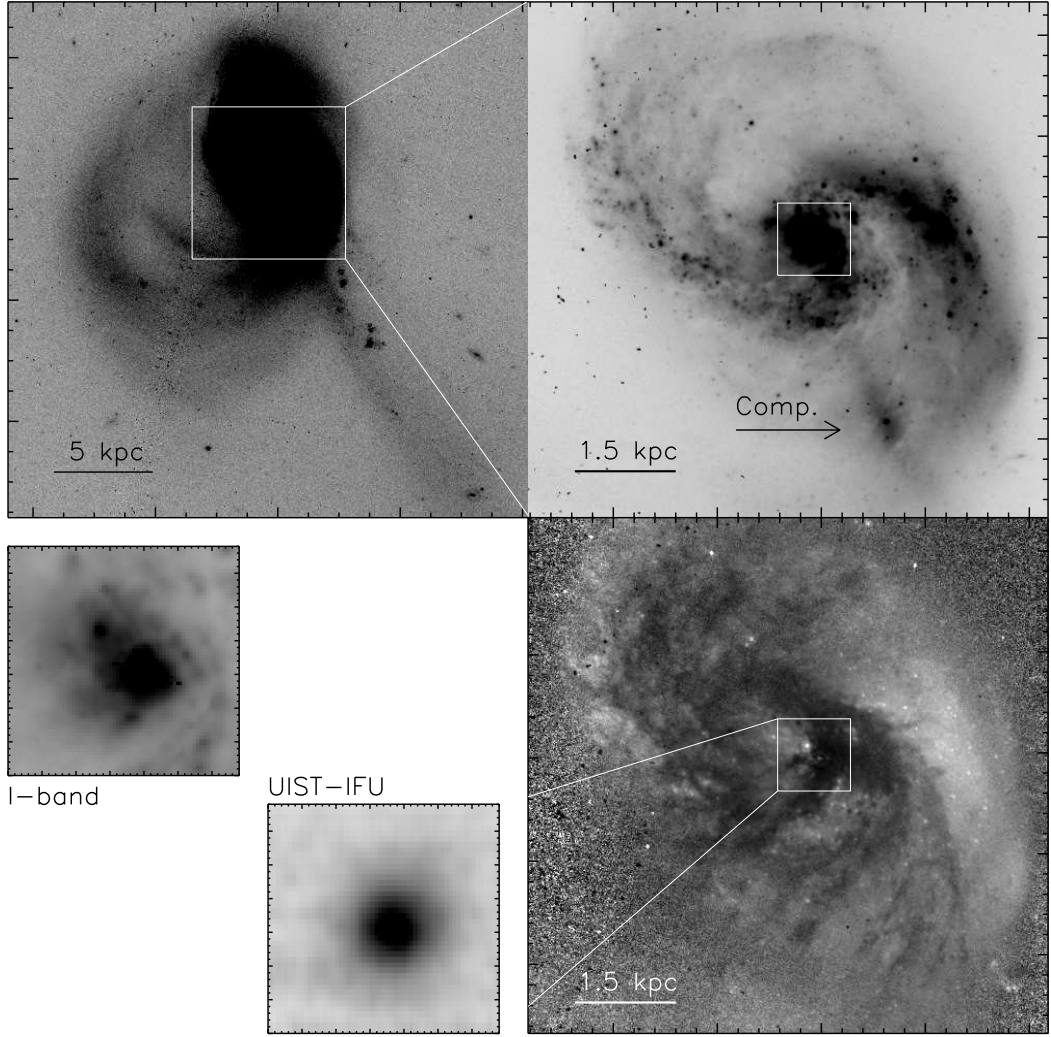}\hspace{2pc}%
\caption{\label{label} \small Archival ACS/HST images of NGC~1614. North is up and 
East left in each image, the tick-marks are 5" in the top-left image
and 1" in the right side panels.  {\em Top-left}:  an $I$-band (FW814) image 
showing the outer structures of NGC~1614 at 20 kpc scales. The white rectangle 
shows the area of the same image zoomed and re-scaled at {\em top-right}, 
showing the more regular spiral pattern. The higher surface density structure
9" SW of the nucleus is pointed out, which is the likely remnant of a companion
galaxy, and possibly the trigger for the starburst core of the whole system.  
The {\em bottom-right} panel shows 
the ACS/HST $B-I$ colour map of the same region as the panel above it, where 
darker areas mean redder colours, i.e.\ higher extinction;  the main nucleus lies behind a 
dust feature.  The innermost 1 kpc area is marked with the white
rectangles in both right-hand-side panels and is shown zoomed-in at  {\em bottom-left}.
The left one is the $I$-band image, and at right we show our UIST
$L$-band continuum image, equivalent of Fig.~2., right panel. The bright very blue point source
$\sim1.5$" NE of the nucleus visible in the I-band and in the colour-map is a likely young 
super star cluster.}
\end{figure}

We detect a strong central continuum  source which is slightly resolved ($\sim$ 80 pc; Fig.~1 lower left, Fig.~2), and a clear signature of a PAH 3.3 $\mu$m ring around the nucleus (Fig.~3).  
Parameterising the 
PAH flux and the L-band continuum following Risaliti et al. (2006) we find no evidence at all for an AGN, obscured or otherwise:  
the continuum is flat, the PAH feature is strong even within the central 30 pc 
resolution element, and there is no 3.4 $\mu$m absorption (Fig.~2).

\begin{figure}[h]
\begin{centering}
\includegraphics[width=17.5pc]{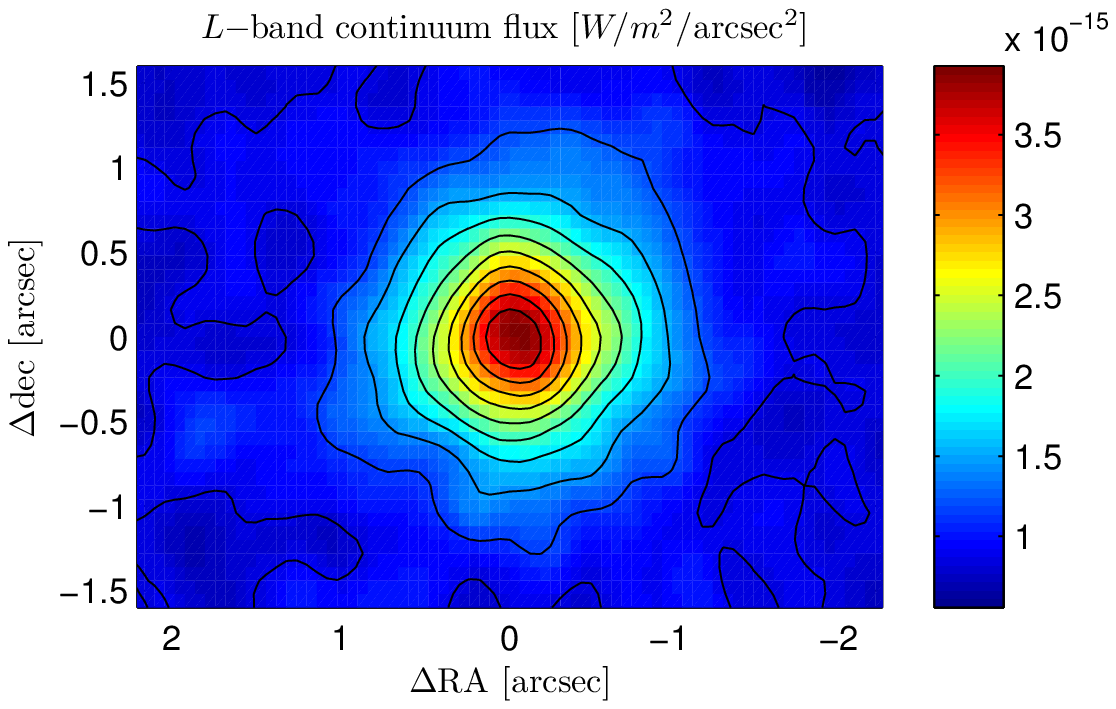}
\includegraphics[width=16pc]{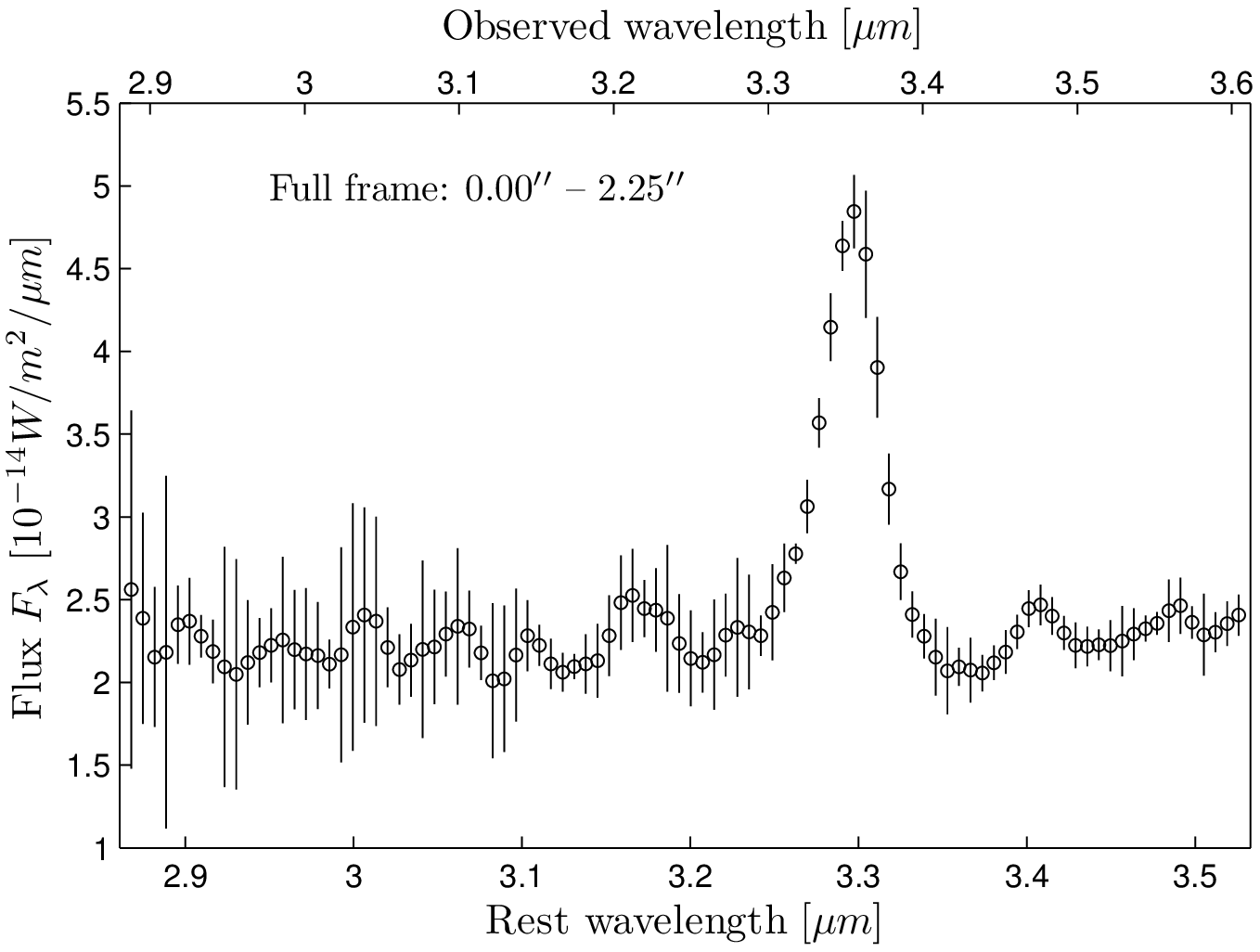}
\caption{\label{label} \small {\em Left:} $L$-band continuum map of the NGC~1614 nuclear regions from the IFU observations showing the strong marginally resolved nuclear source. {\em Right:} L-band spectrum of the central $\sim$ 700 pc area showing a flat continuum and an EW $\approx$ 70 nm PAH feature, typical in starbursts. The spectrum is similar within 0.3" of the core as well.}
\end{centering}
\end{figure}

\begin{figure} 
\begin{centering}
\includegraphics[width=17pc]{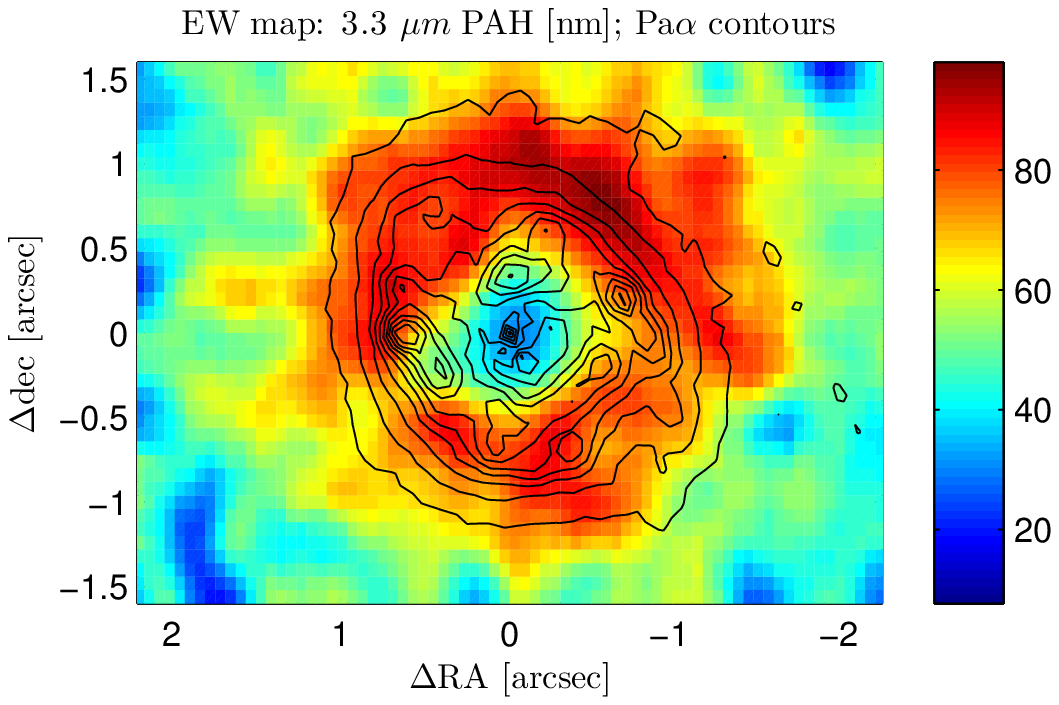}
\includegraphics[width=17pc]{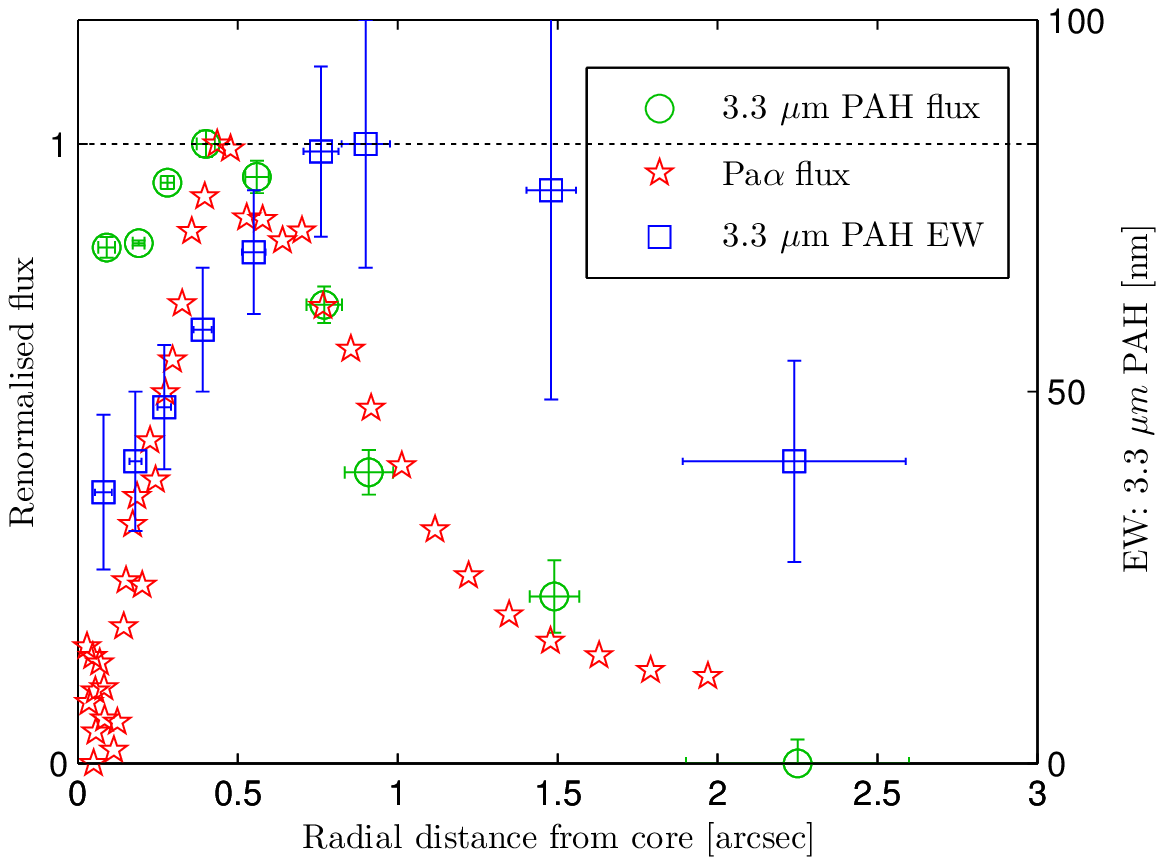}
\hspace{2pc}%
\caption{\label{label} \small {\em Left:}  The NGC~1614 3.3 $\mu$m PAH emission EW shows a wide ring around the nucleus; the contours show Pa$\alpha$ emission{\em Right:} The EW of the 3.3 PAH (blue squares), the PAH feature flux (green circles), and Pa$\alpha$ flux (red stars) shown as a function of radius. Each profile is normalised to unity, though  the actual values of the 3.3 PAH EW can be read from the right-side axis. PAH EW  clearly peaks outward of the main strong SF ring as defined by recombination line strengths. }
\end{centering}
\end{figure}

The PAH ring is especially clear when inspected in terms of its equivalent width (EW) and more or less coincides with, but also extends well outside of, the star-formation ring detected with recombination-lines and radio continuum.  Figure~3 shows the PAH distribution compared to the SF ring measured from archival NICMOS Pa$\alpha$ data.   
We note that the PAH flux extends  to the core regions, where Pa$\alpha$ is not seen, while the PAH
EW strength simultaneously extends quite far outside of the star formation ring.  
Using optical and NIR colour
information of the core, and the  PAH EW strengths in the various regions, and  comparing to
previous studies linking PAH strengths with ages of SF regions (Tacconi-Garman et al. 2005; 
D\'iaz-Santos et al. 2008 ), we find the 
central kpc of NGC~1614 to be consistent with the following scenario:  the core has experienced a 
starburst some 20 to 100 Myr ago, has no more ongoing SF and much of the initial ISM has evacuated 
the region.  The ring at 300 pc radius is the site of strongest current SF, and the age of the SF is 
continuously decreasing as one moves outwards from the core to mid-ring, judging by the change 
of the PAH-to-Pa$\alpha$ ratio.   This ratio grows again outward of the ring, but there the level of
star-formation is likely (still) too low to be detected in recombination lines, while the strong
PAH EW might instead be a signal of  the very  youngest SF (2-4 Myr time scale; Beir{\~a}o et al 2009).
 It thus appears that NGC~1614 is experiencing an outward propagating starburst, the detected PAH
ring highlighting the location where the SF is just propagating into the "new" molecular gas clouds.

Moreover, we detect a likely remnant of the companion galaxy  and likely trigger of the LIRG
phenomenon in the linear tidal tail extending SSW of the nucleus, $\sim$9" away.  Comparing 
to dynamical models (Johansson et al. 2009), this feature is consistent with a remnant of a smaller
spiral galaxy which is now falling towards the main component for the second time, and has likely lost much of its mass, and perhaps is the source of the whole linear tidal tail.  The timescales would put the first passage somewhere at 50 Myr in the past, consistent with the estimates of the age of the central starburst.

\section{A triple merger with a hidden companion}

V\"ais\"anen et al.\ (2008) presented NIR AO-imaging and optical spectroscopy of the system 
IRAS19115-2124.  The galaxy, which we dubbed the Bird, was expected to consist of a 
pair-interaction based on optical HST imaging.  The VLT/NACO imaging and SALT spectroscopy, however, suggested a third component.  (U)LIRGs are complicated dusty objects, and optical data alone, even 
of high spatial resolution, often can miss whole nuclei and major components of the interaction (Fig.~4.; see also Haan et al. 2011).  
Moreover, there was evidence that this third, least massive and
irregular component dominates the current star-formation output of the whole
system.  A similar case was recently presented by Inami et al. (2010). 
This is in contrast to the widely held picture that tidal interactions
are expected to drive large quantities of gas into the {\em central regions} 
of the interaction resulting in nuclear starbursts.   We have now observed the Bird system in more 
detail using MIR imaging in the 11 $\mu$m band with VLT/VISIR and JK IFU observations
with VLT/SINFONI.

\begin{figure}[t]
\includegraphics[width=17pc]{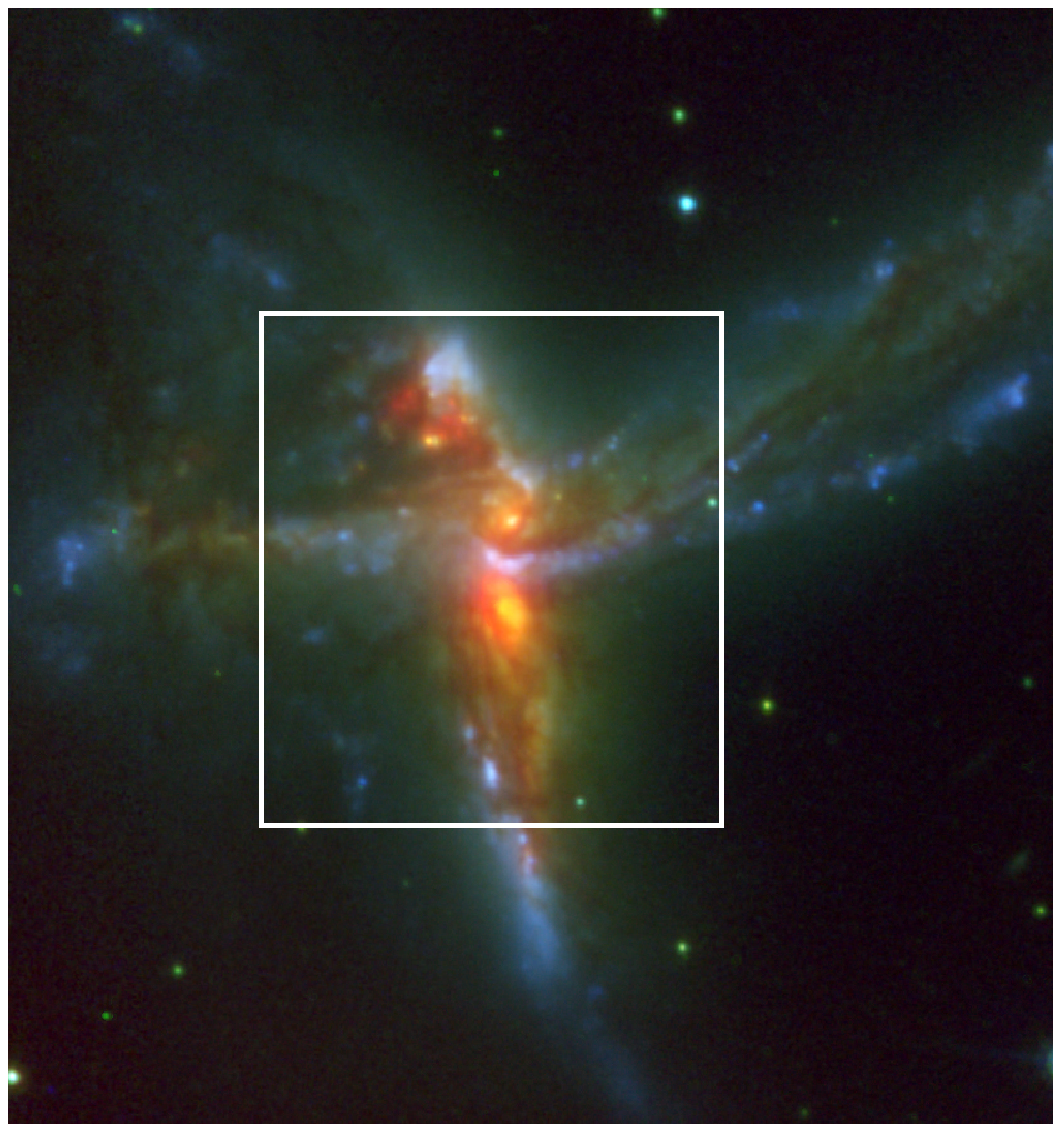}\hspace{2pc}%
\includegraphics[width=16.6pc]{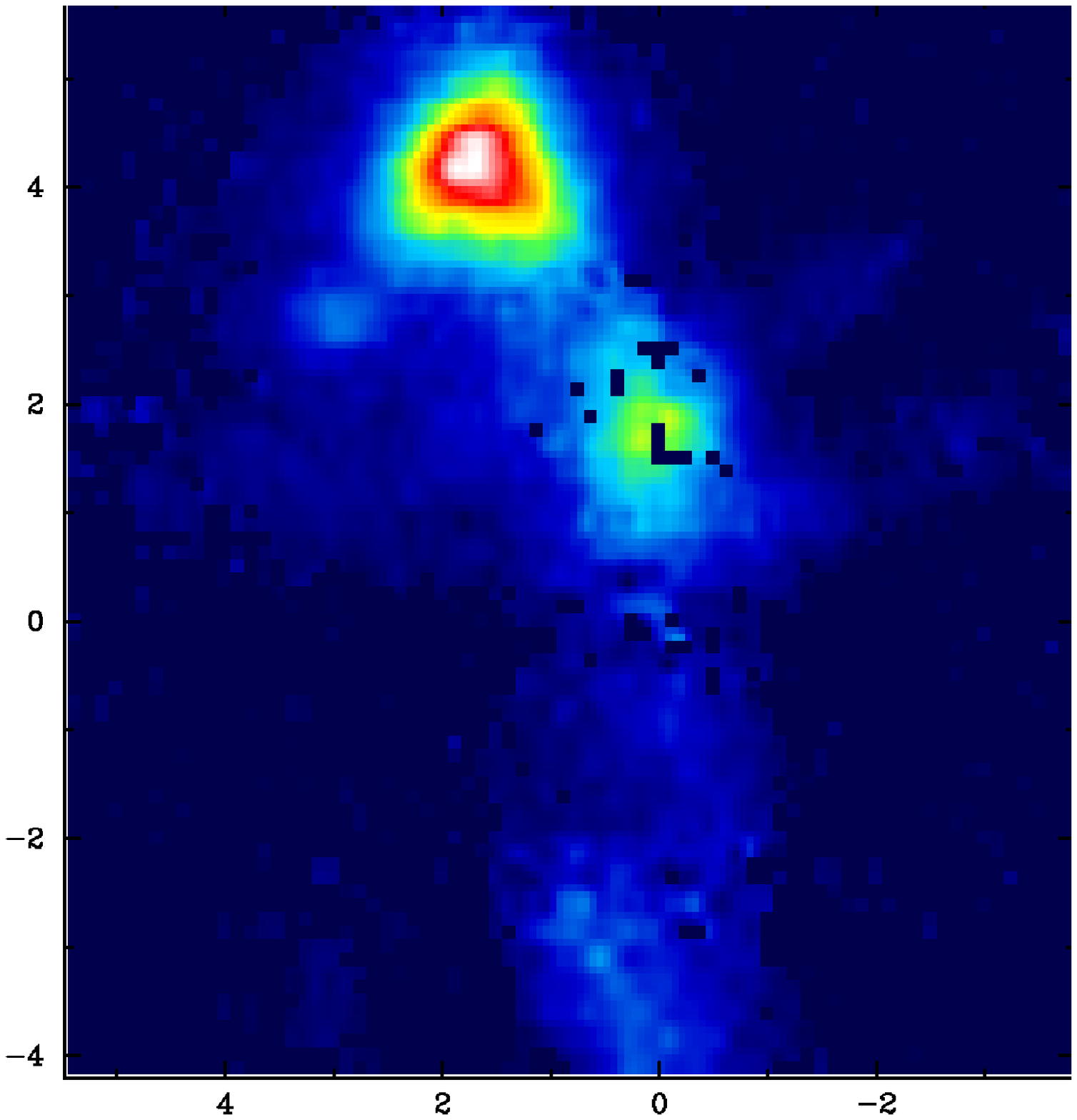}\hspace{2pc}%
\caption{\label{label} \small  {\it Left}:  A BIK-three colour image of the Bird, where optical bands are from HST and the K-band is our VLT/NACO AO-data.  Two major nuclei are seen in the centre.  While the optical HST images show spectacular tidal tails extending both vertically and horizontally, the larger Southern nucleus  ("Body") is virtually invisible in the HST B-band image, and also the third component ("Head" of the Bird) North-East of the two central nuclei is also totally inconspicuous.  The white box shows the region observed with VLT/SINFONI.  {\it Right}:  The SINFONI Pa$\beta$ map of the Bird.  The current SF of the Bird is dominated by the small and irregular "Head" component, which has an estimated mass of 1/10 of the two other components.  There is  some SF also happening in the
smaller of the major nuclei ("Heart"), which still resembles a spiral galaxy in the K-band images regardless of the fairly advanced interaction.  The tick marks are in arcsec, and 1 arcsec corresponds to approximately 1 kpc.}
\end{figure}

Figure 4., right panel, shows the Pa$\beta$ flux over the Bird system, which clearly locates the current SF into the "Head" component.   About 3/4 of the current SFR originates from this minor component.  
While the analysis of the whole dataset is still ongoing, several physical characteristics are immediately obvious from the SINFONI IFU spectra.  The Br$\gamma$ EW is an order of magnitude larger in the "Head" than in the "Body", and mid-way in the "Heart", indicating youngest SF in the Head.  Interestingly the [FeII]/Pa$\beta$ ratio in the Head component is larger in its outer parts  compared to its inner regions, suggesting the current SF is more compact than the somewhat older SF of few 10s of Myr on the outer rim.  Ratios of molecular hydrogen to hydrogen recombination lines show LINER-like values over the Body component while being pure starburst-like elsewhere.  

Figure~5.\ shows the 11.8 $\mu$m contours from VISIR data, corresponding to 11.2 $\mu$m PAH-feature at z=0.049, overlaid on the K-band NACO data. The strongest PAH emission is coming from the "spiral galaxy" at the Heart of the Bird.  While there is significant PAH flux from the Head as well, it does appear that the flux is depleted there, possibly by the destruction of PAH carriers or e.g.\ by dilution of hotter dust continuum.  The Spitzer/IRAC $3.6 - 4.5 \ \mu$m colour of the Head, in fact, is very red, typical to dwarf galaxies (e.g. Smith et al. 2009).

Some example velocity slices extracted from the Pa$\alpha$ and H$_2$ lines are shown in Fig.~6 to show the quality of the data and the main kinematic components. The velocity of the Head component is quite much higher than the others and actually extends to 15200 km/s.  When examining the full data-cube it appears that there are ionised gas {\em outflows} from both the two main galaxy nuclei -- cool gas outflows were already detected by the optical SALT spectroscopy in NaD absorption. In addition, there are also indications of {\em inflows} of gas at the Body and Head components.

\begin{figure}[t]
\includegraphics[width=14pc]{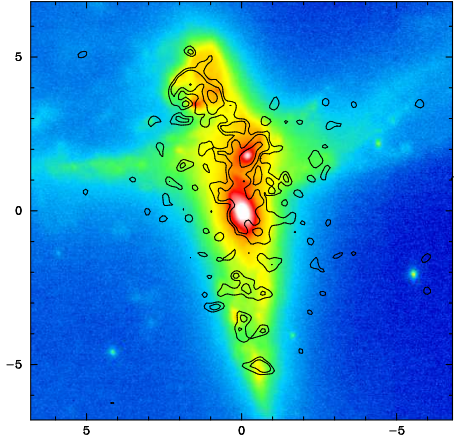}\hspace{2pc}%
\begin{minipage}[b]{16pc}\caption{\label{label} \small VLT/VISIR 11.8 $\mu$m contours on the K-band image.  There appears to be a depletion of PAH strength at the location of the hottest component, the Head of the Bird.}
\end{minipage}
\end{figure}

\begin{figure}[t]
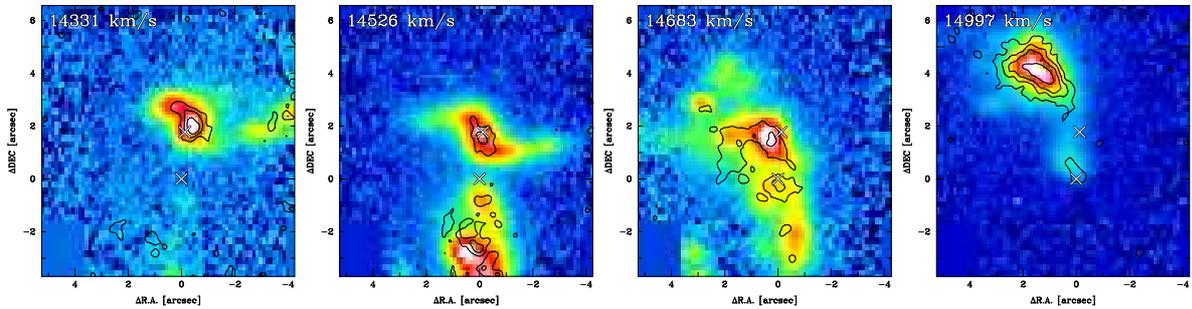

\includegraphics[width=9pc]{plot1.eps}
\includegraphics[width=9pc]{plot2.eps}
\includegraphics[width=9pc]{plot3.eps}
\includegraphics[width=9pc]{plot4.eps} \hspace{4pc}%
\caption{\label{label} \small Pa$\alpha$ maps at selected velocities from our SINFONI data, with the corresponding H$_2$  emission shown as contours.  The latter follow the Pa$\alpha$ distribution very well over the Head component, but the distributions are more varied over the main nuclei. The positions of the two main K-band nuclei seen in Fig.~5 above, are marked with crosses.  }
\end{figure}

\section{SNe in the nuclei of LIRGs}

SF dominated (U)LIRGs are expected to hide in their central regions large
numbers of undetected core-collapse supernovae (CCSNe), i.e.\ stars more
massive than $\sim$8~M$_{\odot}$ exploding at the end of their (short) lives. Such SNe cannot be detected  at optical wavelengths, even in the local universe, because of severe ($A_V > 10$ mag) dust extinction. Recently, Horiuchi et al. (2011) indicated that the measured cosmic CCSN rate is a factor $\sim$2 smaller than that predicted from the measured cosmic SFR, suggesting there is a "SN rate problem". Therefore,
an accurate correction for the number of SNe lost in starbursts, LIRGs and ULIRGs will be crucial for the present and future CCSN surveys at high-$z$.

Our SN search using Gemini/ALTAIR/NIRI has so far produced the detections of 6 SNe. The most recent ones, SNe 2010cu and 2011hi, both occurred in the same host galaxy IC 883 and were located at only $\sim$200-400 pc projected galacto-centric distances.  Both the SNe seem to suffer from relatively low host galaxy extinction $(A_V < 1$ and $< 7$, respectively; Kankare et al. 2012) suggesting they are not deeply buried in the nuclear regions of IC 883. Despite the rather low extinctions such SNe occurring close to the LIRG nuclei have remained undiscovered by the current SN searches and will therefore contribute to the "SN rate problem". Moreover, some of the other SNe we have discovered have had significantly higher extinctions (e.g. Kankare et al. 2008).

Based on the IR-luminosities for our sample of 7 LIRGs in the multi-epoch Gemini observations, the intrinsic number of CCSNe exploding over the program is about 15 CCSNe. We expect about one third of these to be detectable depending on the assumed distribution of extinctions and concentration of SNe within the innermost nuclear regions. Although the number statistics are arguably still low and the nature of (U)LIRGs at higher redshifts not precisely known, it already appears clear that a significant fraction of CCSNe at intermediate and high-$z$ remains undetected in these systems. Therefore,
AO-assisted NIR observations provide an important window to detect and study SNe in the obscured nuclear regions of the local counterparts of
the galaxies dominating the SF at higher-$z$.

\section{Super Star Clusters}

\begin{figure}[t]
\centering
    \begin{tabular}{c}
  {\resizebox{53mm}{!}{\includegraphics{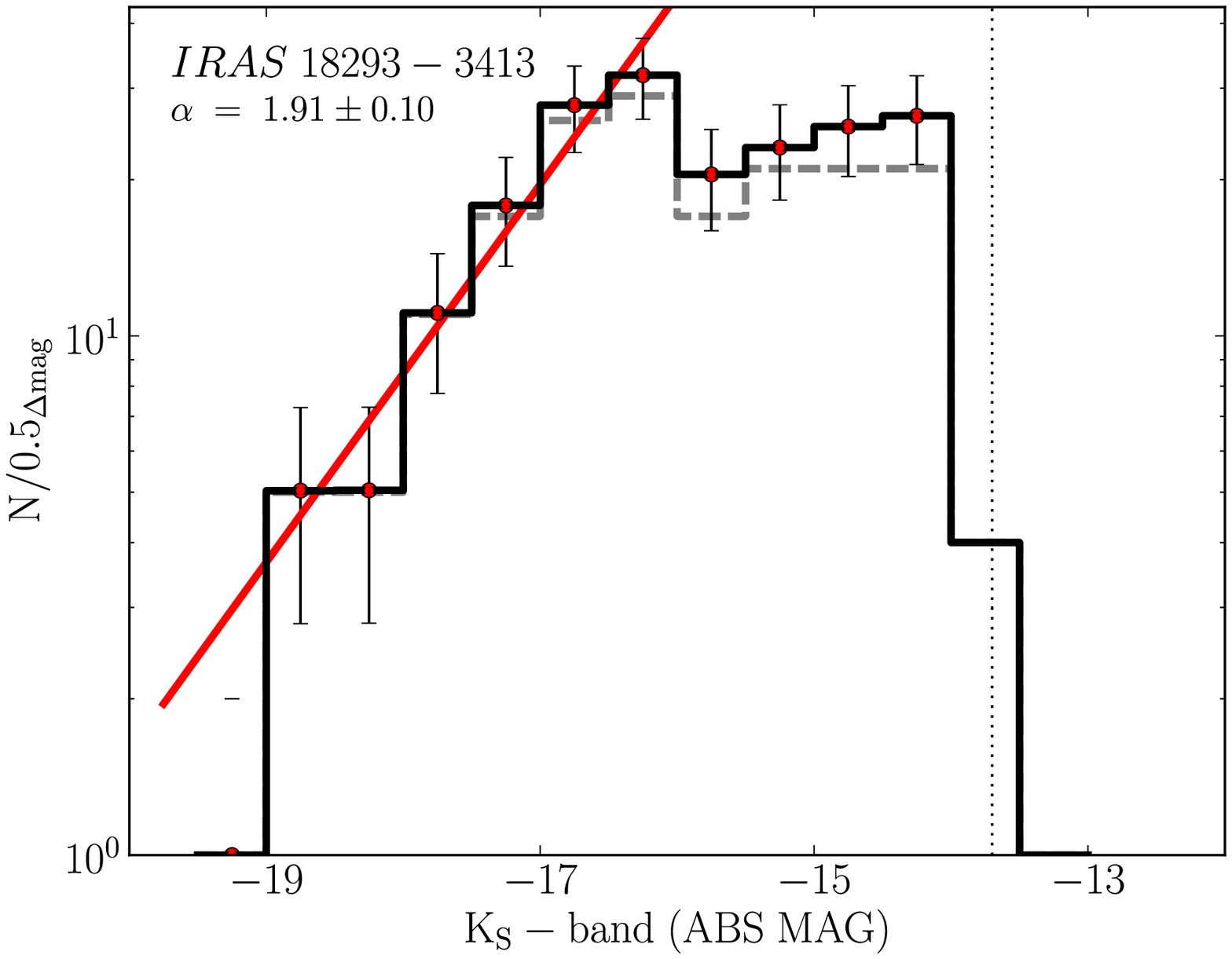}}}
  {\resizebox{53mm}{!}{\includegraphics{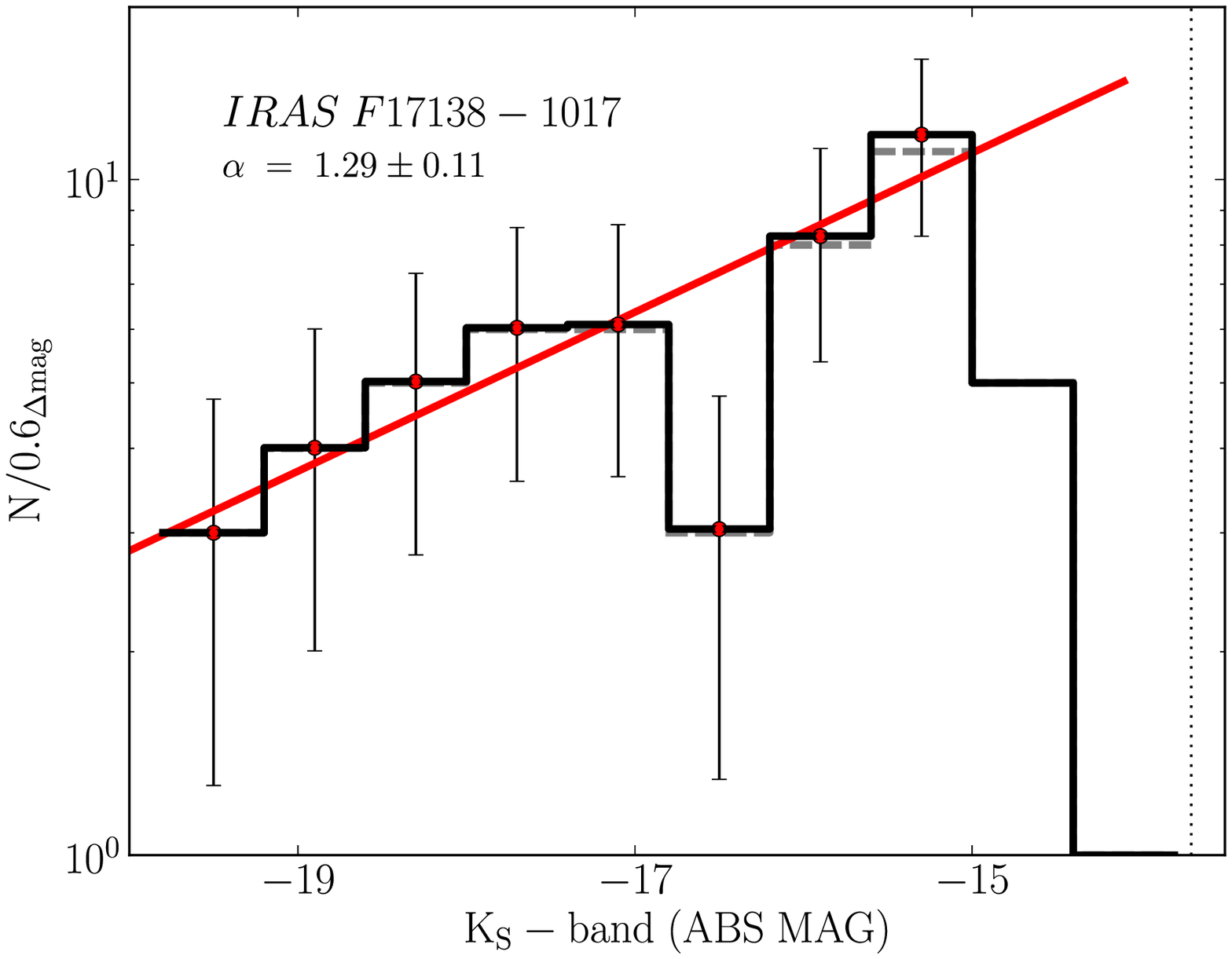}}}
  {\resizebox{53mm}{!}{\includegraphics{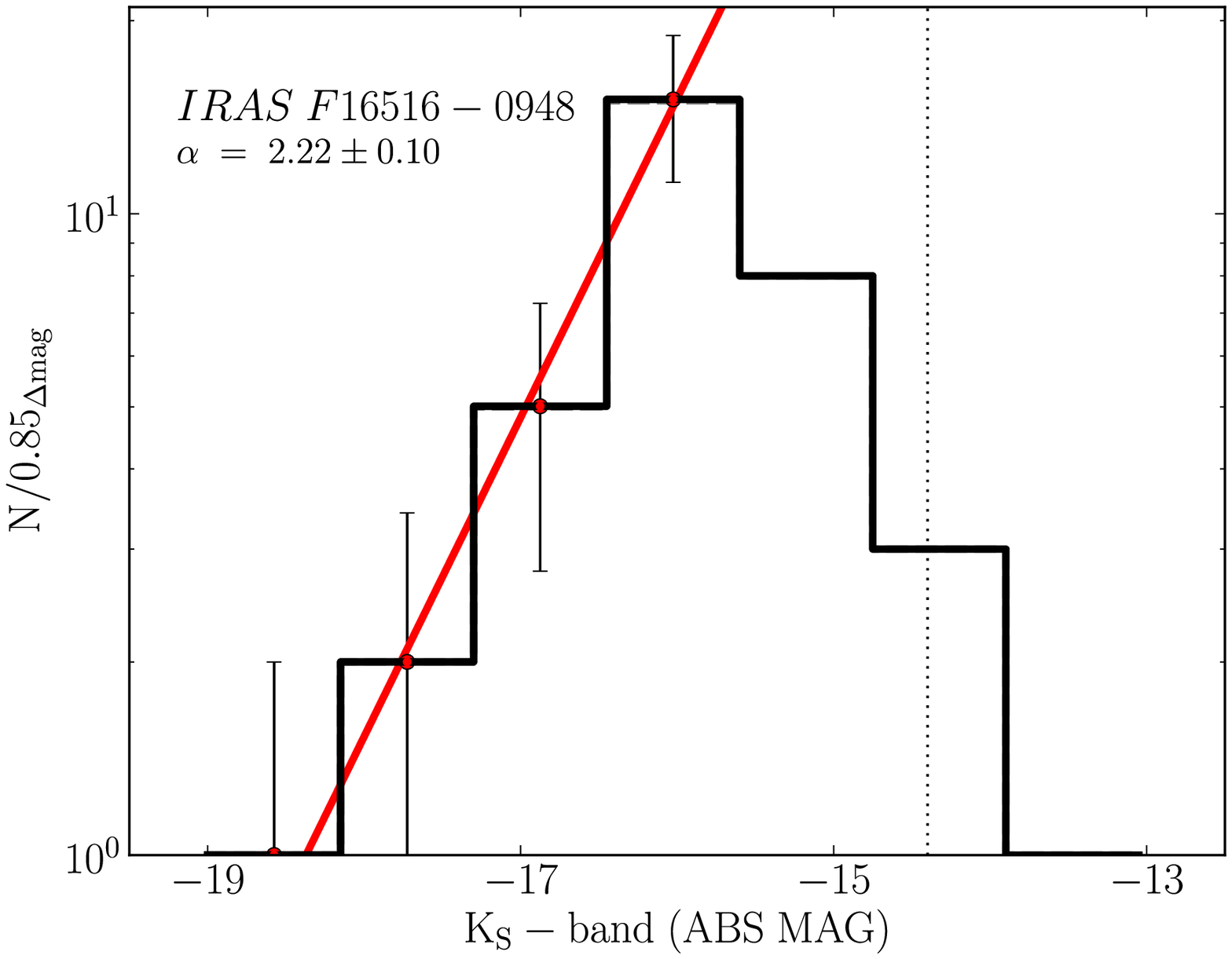}}}
    \end{tabular}
\caption{\small Examples of K-band luminosity functions of three galaxies.  A clear break in the power-law slope is seen on the left, while the other two examples show shallow 
($\alpha \sim -1.3$) and steep $\alpha \sim -2.2$ slopes, respectively, with a 
possible two-component power-law in the distribution on the right as well.}
\end{figure}

Young and massive super star clusters (SSCs) are found whenever very active star formation is going on, such as that in interacting LIRGs (e.g.\ Whitmore et al. 1995). We are studying the luminosity and mass functions of SSCs as a function of SFR and type of host galaxy, interaction stage, etc., in an effort to constrain the cluster mass function (CMF), study the formation, evolution, and disruption of SSCs, which all have 
relevance also to star formation laws in general (Escala \& Larson 2008).  As a first step, we have studied the K-band LFs of the sample of galaxies from our Gemini observations.  According to theoretical expectations and many optical studies of nearby star-forming galaxies, SSC LFs are well fitted by a single power-law index of $\alpha \sim -2$ in a  $\Phi(L) dL \propto L^{-\alpha} dL$ distribution.  We, in contrast, find a surprising variety of slopes, ranging from much shallower ones of $\alpha \sim -1.3$, to more or less the "nominal" $-2$ case.  In addition, intriguingly, most of the galaxies in the initial sample have clear break-points in their SSC LF slope, where the bright section is significantly steeper than the faint magnitude ranges, or we find distributions which might be better fit with e.g. Gaussian distributions.   Fig.~7 shows three examples of the LFs, one with a clear break-point, and others with the opposite ends of the range of slopes, with
$\alpha \sim -1.3$ and $-2.2$, respectively, as fitted to the bright part of the LF.  These effects persist through rigorous photometric, blending, and detection/completeness analyses and simulations in the complex background conditions of the target galaxies.

A range of LF slopes and/or breaks in the slopes could indicate mass-dependent disruption of the SSCs, or environment and/or host SFR dependent cluster CMFs (e.g. Bastian 2008; Chandar et al. 2010; Larsen 2009).  Most previous determinations are from less luminous IR-galaxies, while recently  relatively shallow slopes for SSC LFs have been found in blue compact dwarfs (e.g.\ Adamo et al. 2011 and references therein).  Our analysis is still ongoing, and will be presented in Randriamanakoto et al. (2012, in preparation), and the full sample of 40 galaxies will also be added to the SSC analysis to get a handle on how universal CFLs are, and how they depend on host galaxy properties.

 \section{Summary}
 
 We are studying a sample of several dozen strongly star-forming galaxies, mostly LIRGs, using NIR adaptive optics imaging, and optical spectroscopy, and IFU observations for selected targets.  Adaptive optics and near- and mid-IR observations are often necessary to uncover how many nuclei are involved in the interactions, and to characterise the components, and how star formation is triggered and is propagating in the systems.  We find the L-band, and in particular the 3.3 $\mu$m PAH emission, to trace well various levels of star formation in a LIRG nucleus.  We also find hundreds of super star clusters, and detect a population of  dust-obscured core-collapse SNe in our LIRGs.




\smallskip

\end{document}